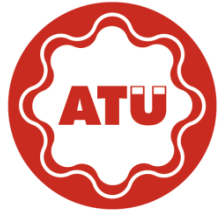

REPUBLIC OF TURKEY

ADANA ALPARSLAN TÜRKEŞ SCIENCE AND TECHNOLOGY UNIVERSITY

GRADUATE SCHOOL OF NATURAL AND APPLIED SCIENCES
DEPARTMENT OF ELECTRIC ELEKTRONIC ENGINEERING

SHARED DATA GRANULARITY: A LATENT DIMENSION OF PRIVACY SCORING OVER ONLINE SOCIAL NETWORKS

YASİR KILIÇ
MASTER OF SCIENCE

ADANA 2020

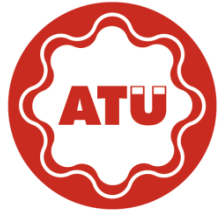

# REPUBLIC OF TURKEY
# ADANA ALPARSLAN TÜRKEŞ SCIENCE AND TECHNOLOGY UNIVERSITY

# GRADUATE SCHOOL OF NATURAL AND APPLIED SCIENCES
# DEPARTMENT OF ELECTRIC ELEKTRONIC ENGINEERING

# SHARED DATA GRANULARITY: A LATENT DIMENSION OF PRIVACY SCORING OVER ONLINE SOCIAL NETWORKS

## YASİR KILIÇ
## MASTER OF SCIENCE

### SUPERVISOR
### Asst. Prof. Dr. ALİ İNAN

**ADANA 2020**

I hereby declare that all information in this thesis has been obtained and presented in accordance with academic rules and ethical conduct. I also declare that, as required by these rules and conduct, I have fully cited and referenced all information that is not original to this work.

[Signature]

Yasir KILIÇ

# ABSTRACT

**SHARED DATA GRANULARITY: A LATENT DIMENSION OF PRIVACY SCORING OVER ONLINE SOCIAL NETWORKS**

Yasir KILIÇ

Department of Electric Electronic Engineering

Supervisor: Asst. Prof. Dr. Ali İNAN

January 2020, 48 pages

Privacy scoring aims at measuring the privacy violation risk of a user over an online social network (OSN). Existing work in the field rely on possibly biased or emotional survey data and focus only on personel purpose OSNs like Facebook. In contrast to existing work, in this thesis, we work with real-world OSN data collected from LinkedIn, the most popular professional-purpose OSN (ProOSN). Towards this end, we developed an extensive crawler to collect all relevant profile data of 5,389 LinkedIn users, modelled these data using both relational and graph databases and quantitatively analyzed all privacy risk scoring methods in the literature. Additionally, we propose a novel scoring method that consider the granularity of data an OSN user shares on her profile page. Extensive experimental evaluation of existing and proposed scoring methods indicates the effectiveness of the proposed solution.

# ÖZET

PAYLAŞILAN VERİNİN ÖGE BOYU: ÇEVRİMİÇİ SOSYAL AĞLARDA
GİZLİLİK SKORLAMANIN ÖRTÜK KALMIŞ BİR BOYUTU

Yasir KILIÇ

Elektrik Elektronik Mühendisliği Anabilim Dalı

Danışman: Dr. Öğr. Üyesi Ali İNAN

Ocak 2020, 48 sayfa

Gizlilik skorlaması bir ÇSA üzerinde kullanıcıların paylaşımlarına göre gizlilik ihlali riskini ölçmeyi amaçlar. Bu alanda yapılan çalışmalar muhtemel ön yargılı ve duygusal anket verilerine dayanır ve yalnız Facebook gibi kişisel amaçlı ÇSA'lara odaklanır. Mevcut çalışmaların aksine, bu tezde en popular ProÇSA olan LinkedIn'den toplanan gerçek dünya ÇSA verisiyle çalışıyoruz. Biz 5,389 LinkedIn kullanıcısının ilgili profil bilgilerini toplamak için kapsamlı bir gezgin programı geliştirdik, toplanan verileri hem ilişkisel hem de çizge veritabanında modelledik ve literatürde yer alan Gizlilik risk skorlama methodlarını niceliksel analiz ettik. Ayrıca, ÇSA kullanıcıların profil sayfası üzerindeki verilerinin öge boyunu düşünen yeni bir skorlama metodu öneriyoruz. Var olan ve önerilen skorlama metodlarının kapsamlı deneysel değerlendirilmesi bu çözümün etkinliğini gösterir.

**Anahtar Kelimeler:** gizlilik, çevrimiçi sosyal ağ (ÇSA), gizlilik tutumu, LinkedIn, öge boyu, profesyonel amaçlı çevrimiçi sosyal ağ (ProÇSA), madde cevap teorisi (MCT)

*I sincerely dedicate this thesis to my family who supported me everytime …*

# ACKNOWLEDGEMENTS

I would like to thank my family who give me pyschological support every time and my supervisor who provided me with superb knowledge and his experience as wonderful supervisor.

# TABLE OF CONTENTS

# LIST OF FIGURES

# LIST OF TABLES

# NOMENCLATURE

| | |
|---|---|
| PS | Privacy Score |
| $\beta_i$ | Sensitivity of profile item $i$ |
| $\alpha_i$ | Discrimination of profile item $i$ |
| OSN | Online Social Network |
| ProOSN | Professional Purpose Online Social Network |
| PerOSN | Personal Purpose Online Social Network |
| PSN | Policy Based Privacy Score assigned by the Naïve Model |
| IRT | Item Response Theory |
| PSI | Policy Based Privacy Score assigned by the IRT Model |
| PSGN | Granularity Based Privacy Score assigned by the Naïve Model |
| PSGI | Granularity Based Privacy Score assigned by the IRT Model |
| PSC | Centrality Based Privacy Score |
| PSNA | Network Aware Privacy Score |
| GM | Granularity Matrix |
| PM | Privacy Policy Matrix |
| GLM | Granularity Level Matrix |
| GRM | Graded Response Model |
| RDB | Relational Database |
| GDB | Graph Database |

# 1. INTRODUCTION

According to recent statistics, the use of Online Social Network (OSN) services is ever increasing (We Are Social Ltd., 2018). OSNs have turned into a vivid, digital reflection of our real-life activities and social interactions (Wani, Agarwal, Jabin, & Hussai, 2018). Such popularity has created room for specialization in OSN services: Facebook and Google+ are mostly used for personal purposes such as acquiring new friends or connecting with old ones, Twitter is preferred for commenting on public events and news, Instagram is the focal point of sharing pictures and last but not the least, LinkedIn caters for professional goals such as exploring possible job opportunities and strengthening ties with past and present colleagues and peers. Various data leakage incidents, one of the latest being the "Cambridge Analytica Scandal of Facebook" (Confessore, 2018), have raised concerns on how the content posted on these platforms should be maintained towards protecting individual privacy.

The right to privacy is recognized globally by the Universal Declaration of Human Rights. Protection in digital domains is guaranteed by additional layers of protection such as The European Union's General Data Protection Law (GDPR) and Turkey's Regulation of Personal Data Protection. However, these laws and regulations become void when users share consentingly on OSNs. Therefore, it is vital that OSN users be informed of the potential risks of sharing data over OSNs. Notice that every user's risk is unique to herself and depends on the type of data shared, and the audience the data are shared with. Recent works (Domingo-Ferrer, 2010) (Liu & Terzi, 2009) (Petkos, Papadopoulos, & Kompatsiaris, 2015) (Srivastava & Geethakumari, 2013) on the subject tries to quantify the overall risk on a personal basis and calls this a user's privacy risk score. The concept is adopted from credit risk scores of the financial system. Higher privacy risk scores imply being more prone to a privacy violation. In this study, we focus on the problem of privacy scoring over professional OSNs –specifically, the LinkedIn OSN, which is the most prominent, popular example of its kind. There are many reasons why professional OSNs deserve special attention:

- Most users of ProOSNs reveal their true identity. This is a basic necessity for benefiting from the OSN services. Fake, anonymous or pseudonymous accounts are very unlikely.

- Data posted on ProOSNs tend to be more truthful than other OSNs. No company would hire someone whose graduation dates on her LinkedIn profile disagrees with her diplomas.
- The privacy paradox (Barnes, 2006) is at its peak in professional OSNs: as a user shares more, she gets more visible but also more prone to privacy violation. Visibility is key to benefiting from ProOSNs.

Existing work on privacy scoring over OSNs rely on survey data. Each user is given a list of profile items such as mother's maiden name, phone number etc. The users are then asked to specify how visibly they would share the item on their OSN profile: with no one, only friends, friends of friends, or everyone. All analyses are carried out over these survey responses. We argue against this approach and claim that scoring in this manner would produce biased or emotional results (Braunstein, Granka, & Staddon, 2011). Collecting data through surveys forces a user to pick the answer she feels is the right choice. The actual privacy policy setting (i.e. the real behavior) of the same user may be drastically different because (i) available privacy settings do not exactly match the user's preferences, (ii) the user is confused about how the settings operate, or (iii) the user does not care enough about privacy to go beyond the default privacy settings. Also notice that whatever privacy risks a user takes by joining an OSN, those risks will be realized based on her shared data - definitely not her survey responses. In this study, we propose generating privacy risk scores based directly on actual user profiles.

A crawler program starts at a specific LinkedIn account (i.e., the seed) and visits the entire network (i.e., up to and including 2nd degree connections) of the seed in breadth-first traversal order. The crawler is responsible from only discovering profiles to be visited. For each discovered profile, a separate profiler program inspects the corresponding user's profile page. Each user's visible data, alongside her connections are then collected and stored in a graph database. Our data collection not only generates a higher quality, much more realistic view of users' privacy behavior but also facilitates a much larger sample. We profiled a total of 5,389 distinct users. Liu and Terzi surveyed only 153 users (Liu & Terzi, 2010), Pensa and di Blasi surveyed only 101 users (Pensa & Blasi, 2016).

In addition to modifying the data collection process, we also propose incorporating the shared data granularity into the privacy risk scoring method. Very simply put, our claim is that the privacy risk of a user should depend on (i) which sensitive items a user shares (Liu & Terzi, 2009), (ii) network centrality of the corresponding user (Pensa & Blasi, 2016) (Pensa, Blasi, & Bioglio, 2019), and (iii) at what detail an item is shared, indicated by the shared data granularity. The intuition behind is as follows. Both user $u$ and user $v$ might make their education data public. If $u$ shares her education history from primary school through college and $v$ shares only college education, the risk $u$ takes should be measured higher than $v$.

The main contributions of our study are as follows:
1. We study privacy scoring over professional OSNs for the first time. No previous study has focused on professional OSNs.
2. We rely on real world data and not user surveys. Therefore; we experiment with observed (versus alleged) behavior of OSN users.
3. We experiment on a much larger data set compared to existing works.
4. We propose shared data granularity as a latent dimension of privacy scoring over OSNs and show how this dimension can be integrated into existing IRT (Baker, 2004) framework.
5. We empirically show through extensive experiments that shared data granularity does contribute to privacy scoring significantly.

The rest of the thesis is organized as follows. We review related work in the literature in Chapter 2. Used materials and methods are explained in Chapter 3. Chapter 4 gives experimental results and their interpretations. We conclude in Chapter 5. Finally, Chapter 6 provides some recommendations on future works.

## 2. LITERATURE REVIEW

OSNs provide an important data environment opportunity. Privacy scoring studies on OSN data are quite common in the literature. These studies are usually survey-based and focus on PerOSN. On the other hand, problems of both data collection and privacy risk scoring over ProOSNs are yet to be studied. Existing work on privacy over ProOSNs generally examine the privacy settings or policies and analyze the privacy violation attacks. (Manzanares-Lopez, Muñoz-Gea, & Malgosa-Sanahuja, 2014) discussed whether the privacy options available to LinkedIn users were sufficient. It is stated that the existing privacy settings do not meet users' privacy concerns. In the study (Caramujo, 2015, July), the privacy policies of the Facebook and LinkedIn OSNs were analyzed and new policy recommendations with higher privacy awareness were given. In addition, unlike these studies, Yang and his colleagues conducted a privacy analysis by assessing possible attack patterns for nearly 9 million LinkedIn user data they collected (Yang, 2012). Although a large number of accounts have been accessed, (Yang, 2012) has significant shortcomings compared to this study: (i) the social environment of the users and the types of connections have not been taken into account such as connections list, also-viewed relations, endorsement interactions not used, and (ii) graph-based database models are not used to conduct comprehensive social queries.

Another interested sub-problem is measuring the privacy risks of users. Firstly, the studies were implemented based on the privacy policies of OSN users on their profile items such as religious belief and education. The first study on this type of privacy scoring of OSN users is due to Liu and Terzi. Their model is inspired by the Item Response Theory (IRT) (Baker, 2004) of psychology. IRT is applied mainly in educational testing. Every question $q$ has a level of difficulty $\beta_q$ and discrimination $\alpha_q$. Primary goal of IRT is to model whether or not an examinee will correctly answer a given question $q$ based on the personal ability level $\theta$ of $e$, and parameters ($\beta_q$, $\alpha_q$) of the question $q$. Liu and Terzi's privacy scoring approach draws the following analogy between educational testing and OSN usage: OSN users represent examinees and attributes that could be shared on user profiles represent questions. A user is assumed to pick the correct privacy setting for an item based on her ability level and the item's privacy sensitivity and discrimination power. When user decisions are binary (e.g., either share the profile item with everyone or no one – mapping to a true/false question), then the problem is said to be dichotomous. Otherwise, if there are more than 2 possible answers/privacy settings, the problem is said to be polytomous. A user's privacy score is simply set as the cumulative product of her privacy settings multiplied by the corresponding item's sensitivity (Liu & Terzi, 2009).

This initial IRT solution by Liu and Terzi (Liu & Terzi, 2009) has laid the foundation for various others. In (Liu & Terzi, 2010), Liu and Terzi incorporate in their model, the underlying OSN's structure. This is achieved by multiplying each item's visibility by the information propagation factor that represents the fraction of OSN users that are given access to the shared data through the chosen privacy setting. It is also shown in (Liu & Terzi, 2010) that the level of information propagation is negatively correlated with a user's privacy score. The faster any information shared by a user *u* spreads on the OSN, the lower will be this user's privacy score. Srivastava and Geethakumari introduce a measuring approach to privacy scoring for unstructured contents (post, reply, etc.) called Privacy Leakage. The approach extracts sensitive information (e.g, e-mail, birthday, age, etc.) from the non-structured contents. They then calculate the privacy risks according to the sensitivity values of the extracted information (Srivastava & Geethakumari, 2013). Domingo-Ferrer proposes a score called Privacy Risk Functionality which measures how much users share information compared to other users. It has been argued that the privacy risks of a user may vary depending on the sharing of other users (Domingo-Ferrer, 2010). Nepali et al. name their score the Privacy Index. The score calculates how sensitive is the information shared by the users (Nepali & Wang, 2013). Sramka argues that auxiliary information should be incorporated in user scores. This involves locating the evaluated users through simple web searches or joining personal data obtained from several OSNs (Sramka, 2015). Petkos et al. do not only regard the information declared by users for privacy measuring, but also consider the information that can be obtained through inference. They also provide an open source framework for privacy measurement (Petkos, Papadopoulos, & Kompatsiaris, 2015). It is important to emphasize once again that all of these studies. In these policy-based scoring models, granularity is not taken into account for privacy scoring. In other words, these solutions focus only on whether or not an item is shared, but not the granularity in which it is shared. In this study, granularity-based privacy scoring models are proposed.

Another privacy scoring approach is based on users' position on the network. Firstly, it was experimentally demonstrated that centrality can also be a factor for privacy scoring by observing correlation between centralities and policy-based intrinsic privacy scores (Pensa & Blasi, 2016). In another study, it is emphasized that privacy behaviors of neighboring users may affect users (Bioglio & Pensa, 2017). Alemany et al. developed a scoring based on the level of accessibility of users' posts on the OSN, and experimentally demonstrated that this scoring had a strong relationship with the centrality of the users (Alemany, del Val, Alberola, & García-Fornes, 2018). They argue that utilizing the centrality of a user in the network is a good method of generating a privacy score for the same user. As a result, their proposed scoring model disregard an OSN user' profile shares such as education, birthday. In other words, only network topology-based scoring model was proposed. Pensa et al. propose a scoring approach, considering that users' privacy policy-based intrinsic scores as well as those in the social environment affect these privacy scores. According to this approach, if the low risk user has a higher risk social environment, the risk of the user will increase or vice versa (Pensa, Blasi, & Bioglio, 2019). They used survey-based profile items, although they used real network topology. However, in this study, real network topology and real profile information are used for privacy scoring.

# 3. MATERIALS AND METHODS

In this chapter, the data and methods used for the analyzes are explained. The used methodologies for data collection and modelling are detailed in Section 3.1. And then the privacy risk analysis methods applied on the collected data, namely, privacy risk scoring models given in Section 3.2. Our proposed granularity-based privacy scoring approach is presented in Section 3.3. Finally, two statistical methods used for evaluating these different privacy risk scoring models are detailed in Section 3.4.

## 3.1 OSN Data Collection and Modelling

Section 3.1.1 touches on our preliminary data collection efforts and the design choices that resulted from these efforts. The design of the developed automated software for data collection is outlined in section 3.1.2. And finally, we explain how the collected data is modelled in section 3.1.3.

### 3.1.1 Preliminary Efforts for the Data Collection

Before collecting data, we carefully inspected our data collection environment, the LinkedIn OSN and investigated its element structure. The following preferences for the collection process were made:

- In LinkedIn, viewing a member's full OSN profile requires being authenticated with a LinkedIn user account. To overcome this restriction, the credentials of a LinkedIn OSN account were used.
- LinkedIn service provider limits the number of profile searches that can be issued by any member OSN account. Premium accounts are allowed to issue queries. Consequently, we purchased a premium account to speed up the data collection process.
- When the structure of web elements on the data collection environment is examined, it is observed that the environment includes dynamic interactions on the pages such as scrolling to down pages. Javascript-based commands are implemented to automate these interactions on the page.
- OSN User Profile items or attributes that can be obtained from the data collection environment, LinkedIn OSN, have been determined. These determined profile attributes and their entry examples are listed in Table 3.2.

- Python Programming language is preferred because it has some advantages to develop a software program such as wide community support, flexible code writing. And also, the SELENIUM library (Muthukadan, 2019), which can automate user behavior on the browser, is preferred as the software package.
- Two different database models were used to store the collected dataset: Relational Database Model, Graph Database Model. These models are the most popular models used for storing data in literature. MYSQL is used as the underlying relational database management system and Neo4j is as the graph-based database management system.

**Table 3.1** Faced Problems and Their Solutions

| Problem&Constraint | Solution&Preferences |
|---|---|
| Profile Not Viewed without login | A user's credentials |
| Daily Profile Search Limitation | A premium account |
| Dynamic web interactions | Javascript programming |
| Which profile items can be crawled? | See Table 3.1 |
| Which programming language? | Python |
| Which programming library or package? | Selenium package |

### 3.1.2 Data Collection Software Design

In this part, we discuss the design of the web crawler implemented for collecting OSN data. The flow diagram of general operations during data collection is shown in Figure 3.1. First, unique ID information (URL) of the users whose profile information to be collected is obtained by the Crawler module and stored in a MYSQL relational database model. Simultaneously, OSN profiles have been visited with respect to these collected user id information and this was done by Profiler module. This module also stores the collected data in both a relational MYSQL model and a graph-based Neo4j model. These two software program modules, the Crawler and the Profiler, are designed as two distributed partially dependent software modules.

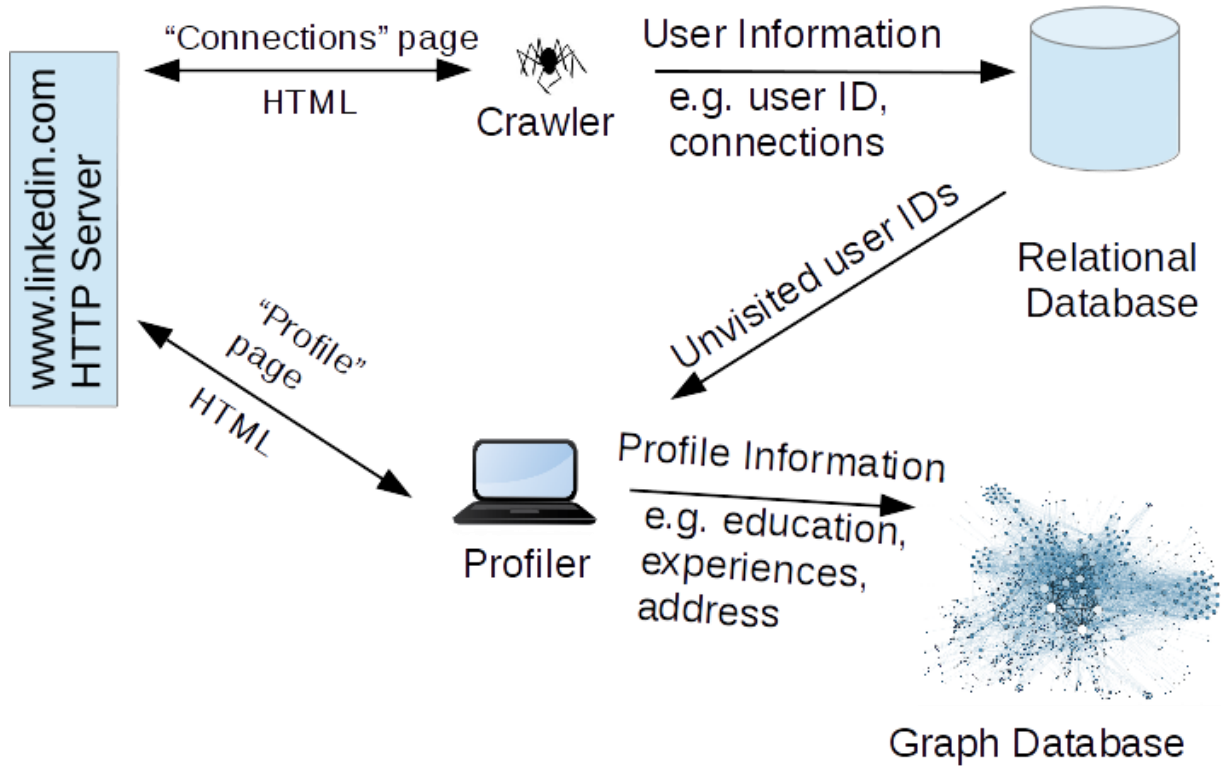

**Figure 3.1** Data Collection Procedures

We developed a web crawler to collect OSN data from LinkedIn by making above design preferences in Table 3.1. Towards this end, we used the Python language and imported the Selenium library (Muthukadan, 2019). Data collected between May 2018 and August 2018 were modelled and queried with the relational MySQL and the graph-based Neo4J database management systems which are explained in detail in section 3.1.3. The crawler visited all directly reachable OSN user profiles rooted at the seed LinkedIn account. In breadth-first traversal order, a total of 5,389 distinct profiles have been accessed. At each profile page, our program recorded every bit of visible data including education history, work experience, phone number, e-mail address and social relations such as connections, endorsements. The complete list of attributes is given in Table 3.2.

**Table 3.2** Collected Dataset Item List

| Item Name | Example Item Entry |
|---|---|
| Birthday | July 20 |
| Connections List | [ainan2435; ykilic7893596] |
| Company Web Pages | [www.examplecompany.com] |
| Personal Web Pages | [www.examplepersonal.com] |
| Phone Number | +90 576 123 45 67 |
| Also Viewed Users List | [ali2wr45; ayşe35446] |
| Education History | [Selcuk University, Computer Eng.,2002,2004] |
| Work Experiences | [Turkcell, Engineer, 2016, 2018,2 years 3 month] |
| User Interest Pages | Steve Jobs; Gandi |
| E-mail | 123example@gmail.com |
| About | I am an engineer at Turkcell |
| Location | Turkey, Adana, Saricam |
| Recommendations | ["Given"," 7893596"," I am definite that every professor would want more students like X in the classroom…"] |
| Endorsements | [ainan2435," Data Mining"] |

### 3.1.3 Dataset Modelling

The collected dataset is stored on databases due to some advantages such as structural querying. Two different types of database models are used: Relational Database and Graph Database models. Generally, this section provides technical details on these two models.

### 3.1.3.1 Relational Databases

This type of databases generally provides storing the data in a tabular format. Data is presented to users as relational tabular format that is collections of tables which have tuples as rows and attributes as columns (Codd, 1970). To manipulate data some operations are also provided to users such as add, delete, update, join on tables. Semantic relationships between tables or entities are applied by key attributes such as primary, secondary key.

Database Management Systems (DBMS) provide an interface for a database model and allowing users to write structural queries on databases using various operators. Relational Database Management Systems (RDBMS) are used for relational database models. Most common language of their structural query is SQL which stands for Structural Query Language ( Wikimedia Foundation, Inc, 2019). This language provides writing structural query on databases.

### 3.1.3.2 Graph Databases

Graph Databases are a database modelling form that uses the graph data structure to model data. In other words, the entities of the data are modelled as nodes and the semantic relationships between these entities as edge ( Wikimedia Foundation, Inc., 2018).

Graph databases uses different indexing mechanism than other databases such as relational database. This mechanism is called Index Free Adjacency (McCreary, 2019). With this mechanism, the data as node is loaded into RAM together with the adjacency physical memory address. This provides a fast memory lookup for queries.

There are many database management systems used for graph databases. In this thesis, Neo4j which is a database management system, is used due to some advantages such as open source, community support, most popular database management system for graph databases (Neo4j, Inc, 2019).

### 3.2 Privacy Risk Scoring Models

A lot of scoring models and their methods have been proposed in the literature to measure the privacy violation risk of OSN users. These scoring models differ according to the calculation approaches and the factors they consider. Most privacy scoring methods in the literature commonly depend on the following factors:

- Sensitivity of profile items: It represents the importance of a profile item is for privacy protection. For example, intuitively, phone number is more sensitive than education history.
- Visibility of profile items: This factor shows visibility levels or sharing probabilities.

- Network position of users: User's network position on the network may affect their privacy. Central users may be more exposed to privacy violation risk.

Existing scoring models can be grouped under the following two categories: privacy policy based and network-based scoring models. Each category will be detailed next.

**3.2.1 Privacy Policy Based Intrinsic Privacy Scoring Models**

OSN users make preferences or settings to protect the privacy of their profile items. These settings define a user's privacy policy on the OSN. This policy information is evaluated in two different perspectives: public view, sharing level. While public view perspective is used to define whether access to the items from public view as dichotomous form which is shown in Figure 3.2, with sharing level perspective is defined which privacy setting or maximum sharing level is used as polytomous form.

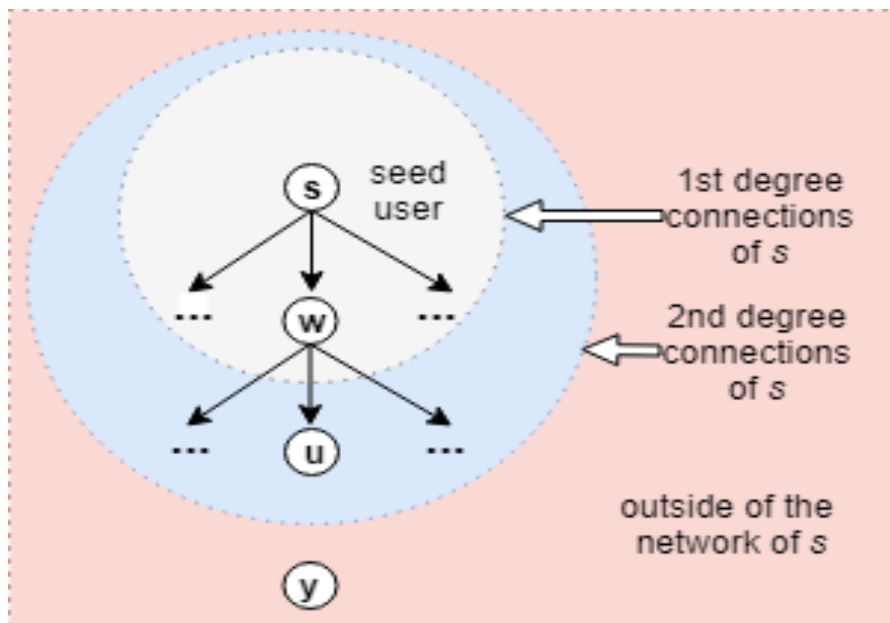

**Figure 3.2** A Social Network Structure with Seed Account Perspective

Generally, most studies about privacy risk scoring on OSNs measure risk by taking into consideration the privacy policy of the users. These scoring models are generally applied in 2 different methods: Naïve, IRT (Item Response Theory) model. Liu and Terzi provide a clear, concise baseline model for privacy scoring called the Naive model (Liu & Terzi, 2010). This model, together with their IRT solution is the basis of almost all studies on the problem of privacy scoring over OSNs, including this thesis. We first introduce this Naive model, then the IRT model. All notation used during this presentation is summarized in Table 3.3. For brevity, we exclude the polytomous problem setting.

**Table 3.3** Terms and their notations

| Terms | Description |
|---|---|
| $n$ | Number of profile items |
| $N$ | Number of users |
| $i$ | Any item |
| $R_{i,j}$ | $n \times N$ Response Matrix |
| $|R_i|$ | Number of 1s for row $i$ of $R$ |
| $|R_j|$ | Number of 1s for column $i$ of $R$ |
| $V_{i,j}$ | User j's visibility on item $i$ |
| $\beta_i$ | Sensitivity of profile item $i$ |
| $P_i$ | Visibility probability of item $i$ |
| $P^j$ | Visibility probability of user j on overall items |
| $PS^j$ | Privacy score of user $j$ |

Policy based intrinsic scoring models are calculated by 2 parameters or factors: sensitivity of item *i* and visibility of item *i*. Once these two factors are calculated, their multiplications give a user's overall privacy risk score according to Equation 3.1.

$$PS^j = \sum_i \beta_i x\, V_{i,j} \qquad (3.1)$$

**3.2.1.1 Naïve Model (PSN)**

This model focuses on privacy policy usage frequencies of users. Users and their privacy policy on profile items are assumed to be independent and scored based on the response matrix. Sensitivity of an item is computed with respect to that item's sharing frequencies. That is, an item is more sensitive if shared by less users. Morover, item visibility is computed with a probabilistic perspective. In this perspective, an item's visibility depends on item visibility probability over all users and user's visibility probability over all items.

In the Naïve model, sensitivity of an item ($\beta_i$) is simply the proportion of users that have hidden away the item in their profiles. Equation 3.2 explains how $\beta_i$ is computed. Intuitively, the more users hide an item $i$, the more sensitive that item is considered to be.

$$\beta_i^{Naive} = \frac{N - |R_i|}{N} \tag{3.2}$$

Visibility indicates the probability that an item of a user can visible to attackers. Let $R_i$ be the $i^{th}$ row of $R$ and $R^j$ be the $j^{th}$ column of $R$. The Naïve model sets visibility($V_{i,j}$) based entirely on $R_i$ and $R^j$ according to Equation 3.3. These probabilities are derived from $R$ by counting number of 1s in the respective row and column. Notice that the two are assumed to be independent.

$$V_{i,j}^{Naive} = P_i \; x \; P^j = \frac{|R_i|}{n} \; x \; \frac{|R^j|}{N} \tag{3.3}$$

The calculated sensitivity values in Equation 3.2 and visibility values in Equation 3.3 for each profile item are multiplied and overall sum gives the user's privacy risk score for Naïve model. The scoring equation of the Naïve Model (PSN) is shown in Equation 3.4.

$$PSN^j = \sum_i \beta_i^{Naive} x\, V_{i,j}^{Naive} \tag{3.4}$$

**3.2.1.2 IRT Model (PSI)**

This scoring approach relies on the Item Response Theory (IRT) which is a psychometric theorem. According to this theorem, the probability of an examinees answering a question correctly depends on three parameters:

- Difficulty of the question,
- Discrimination of the question, and
- Ability of examinees.

This model is inspired by the IRT (Baker, 2004) of psychology. IRT is applied mainly in educational testing. Every question $q$ has a level of difficulty $\beta_q$ and discrimination $\alpha_q$. Primary goal of IRT is to model whether or not an examinee $e$ will correctly answer a given question $q$ based on the personal ability level θ of $e$, and parameters ($\beta_q, \alpha_q$) of question $q$. Liu and Terzi's privacy scoring approach draws the following analogy between educational testing and OSN usage: OSN users represent examinees and attributes that could be shared on user profiles represent questions. A user is assumed to pick the correct privacy setting for an item based on her ability level and the item's privacy sensitivity and discrimination power. When user decisions are binary (e.g., either share the profile item with everyone or no one – mapping to a true/false question), then the problem is said to be dichotomous. Otherwise, if there are more than 2 possible answers/privacy settings, the problem is said to be polytomous. A user's privacy score is simply set as the cumulative product of her privacy settings multiplied by the corresponding item's sensitivity.

A scoring method was proposed to analyze the profile item sharing of users over an OSN by inspiring from IRT (Liu & Terzi, 2010). The IRT model approximates the item characteristics curve that best represents the response matrix $R$ to yield ($a_i, \beta_i$) discrimination and sensitivity of item $i$ and $\theta^j$ ability of user $j$. Then user $j$'s visibility on item $i$ can be calculated according to Equation 3.5.

$$V_{i,j}^{IRT} = \frac{1}{1 + e^{-a_i(\theta^j - \beta_i^{IRT})}} \qquad (3.5)$$

Scoring function is implemented by referring to the parameters of the curve that best fits the privacy policies of OSN users. PSI scoring equation based on obtained paramaters from IRT model is shown in Equation 3.6.

$$PSI^j = \sum_i \beta_i^{IRT} x\ V_{i,j}^{IRT} \qquad (3.6)$$

### 3.2.2 Network Based Scoring Models

In this type of scoring models, scores are generated based on the centrality of users in the network in general and widely. This type of scoring models can be grouped into two models: Centrality Based Privacy Scoring (PSC), Network Aware Privacy Scoring (PSNA).

### 3.2.2.1 Centrality Based Privacy Scoring Model (PSC)

PSC computes privacy score based on the centrality properties of users in the network (Alemany, del Val, Alberola, & García-Fornes, 2018). 3 common types of centrality methods are used: Page Rank Centrality (PRC), Closeness Centrality (CC), Betweenness Centrality (BC).

A statistically strong positive correlation was observed between the scores produced according to the diffusion movements of the OSN users and the central values of the users (Alemany, del Val, Alberola, & García-Fornes, 2018). Accordingly, it is experimentally shown that the centrality value represents the privacy score of the users. In this study, three scoring approaches are described: PRC, BC, CC. We detail each next.

PRC scoring is inspired by the Page Rank algorithm, an approach that sorts web pages in order of importance. This approach calculates the probability that the web pages will be visited based on the references received. PRC scoring takes OSN users instead of web pages, and users' link relationships instead of page transitions or references. In other words, the PRC scores of the users whose connections are high are higher. Equation 3.7 also shows the calculation formula. $\alpha$ and B values are constant to normalize rank values and $k_j$ represents the number of connections of the user. Matrix *A* is adjacency matrix that store neighborhood identity information.

$$PRC^i = \alpha \sum_{j=1}^{|N|} A_{a_j,a_i} \frac{PRC^j}{k_j} + \beta \qquad (3.7)$$

In a social network, if the total distance of a node to other users is low, then the centralization method in which that node is considered more central is Closeness Centrality (Brandes, 2005, February). CC scoring uses this method to measure users' privacy risk on the OSN. Equation 3.8 shows the equation used for this calculation. $|N|$ represents the total number of users, and $d(a_j,a_i)$ represents the distance between users *i* and *j*. It is argued that if a user's total distance to other users is low, the user is more central and has a higher privacy risk.

$$CC^i = \frac{|N|-1}{\sum_{j=1}^{|N|-1} d(a_j,a_i)} \qquad (3.8)$$

In social network theory, the sum of the rates of being the shortest path among other pair nodes, except for one node itself, gives the value of betweenness centrality (Freeman, 1977). A higher value indicates a node on which information flow can be intense and the size of the centrality. The *BC* scoring method uses this metric to calculate the privacy scores of OSN users. It accepts the nodes as OSN users, and the *BC* score is calculated by the rate at which the shortest path between other users, except the users themselves, passes over the calculated user. The equation used in Equation 3.9 is shown. The $\sigma$ function gives the number of shortest paths with respect to given path.

$$BC^i = \sum_{a_j, a_k \in N} \frac{\sigma(a_j, a_k | a_i)}{\sigma(a_j, a_k)} \quad (3.9)$$

### 3.2.2.2 Network Aware Privacy Scoring (PSNA)

The value of OSN users' profile information-based privacy risks as well as the privacy risks of users in their network are important. For example; if a user has a low risk, but if there are high risk users in his neigborhood, the privacy risk of this user is expected to be higher due to his neighbors, or vice versa.

PSNA model is inspired from the Personalized Page Rank Model (Jeh, 2003). The initial weights of web pages are replaced by a policy-based privacy score of users, and a simulated network is generated to produce a combined scoring (Pensa, Blasi, & Bioglio, 2019). Even if a user's privacy score is low, this score increases when that user is connected to users with a comparably higher privacy score. So, this scoring approach tells us that the privacy behavior of users in the user's network will affect the user's privacy risk. Equation 3.10 (Pensa, Blasi, & Bioglio, 2019) shows the scoring function of PSNA model.

$$P = dA^T P + \frac{(1-d)}{\sum_{k=1}^{n} \rho_p^{(V_k)}} \rho \quad (3.10)$$

Since the calculated values are multiplied probability values, they have low values such as 0.000012. Therefore, a normalization operation is applied. Normalized PSNA scoring function is shown in Equation 3.11 (Pensa, Blasi, & Bioglio, 2019).

$$PSNA(V_i) = P(V_i) x \frac{range(\rho_P)}{range(P)} \quad (3.11)$$

## 3.3 A Proposed Privacy Scoring Model: Granularity Based Scoring Models

In addition to privacy policy of OSN users on profile items, granularity that they share can be considered as a factor affecting their privacy violation risk. In this study, it is experimentally demonstrated that granularity is a necessary factor for privacy scoring. In this part, the proposed scoring model that considers not only user's privacy policy on profile items but also user's granularity level on these items is explained. We explain our scoring approach with the example in Table 3.4. The table gives an example of shared data of 4 users over 3 different profile items: Education History, Work Experiences, About. Although these entries are real, they do not belong to these users. Some highlighted motivations of our granularity-based scoring method are explained. For example, it is seen that users share information in different details for education history item. However, existing scoring methods do not take into account the granularity of the shared information. The scoring method proposed in this study computes privacy scores by using the granularity of shared data. The granularity value is evaluated as byte value. GM in Figure 3.3 is the matrix created over the sample given in Table 3.4. Each cell value in this matrix refers to the granularity of the shared data of the users as byte.

Each information that users share on OSNs, especially in free text items such as education history, about, work experiences, may lead to more privacy violation risk. Our scoring approach focuses on this point. It proposes a privacy scoring approach based on level of shared data granularity. For example; when we look at Table 3.4, it is seen that Omer and Fatma share their education history but have different granularity levels. Fatma's sharing of more data indicates that she gets more privacy risk scores from this approach. In contrast to this approach; according to policy-based privacy scoring approaches, these two users have same privacy risk score for education history item. We consider that this is a contradiction. To overcome the contradiction, we propose granularity-based privacy scoring approach.

**Table 3.4** A Snapshot of Users' Shared Data

| User | Shared Data |
|---|---|
| Ali | **Education History**: < Yildiz Technical University, Bachelor, Software Engineering,2010,2014 > <br> **Work Experiences**: <<Kuveyt Turk Participation Bank, Sofware Developer,2011,2012><Mobilion, Android Developer ,2014,2018>> <br> **About**: < I was born in Samsun and originally from Giresun > |
| Ayse | **Education History:** <> <br> **Work Experiences**: < <CNN Turk, Documentary Specialist,2010,2012><TRT, Documentary Specialist,2012,2014><Adobe,GraphicDesigner,2014,2016><Youtube, Youtuber,2016,Current>> <br> **About**:< I am passionately in love with telling stories with words voices images music and videos. I created documentaries about people culture history and environment for CNNTurk TRT and AlJazeera. I have a creative media agency that helps people and brands for discovering and telling their own stories: http://abak.us I gave hundreds of speechs seminars and presentations about business marketing design and technology> |
| Fatma | **Education History:** :< <Derekoy Primary School, Primary School,2000,2005,Samsun><19 Mayıs Anatolian High School, Science,2008,2011><Selcuk University, Bachelor, Computer Engineering,2012,2016><Alpaslan Turkes Science Tech University, Master, Electric Elektronic Engineering,2017,Current>> <br> **Work Experiences:**< Turkcell Ltd Şti, Intern,2015,2016 > <br> **About:** <> |
| Omer | **EducationHistory:**<<BilgiUniversitesi,Bachelor,ComputerScience,2008,2011,Istanbul><Cornell University,Master,Artifiicial Intelligence,2016,2018>> <br> **Work Experiences:** <> <br> **About:** < I was born in İstanbul Turkey at 21 November 1990. I graduated from Bilgi University Computer Science Department. I was in prep school in my first year. I was a candidate for student consul at one point. I attended several seminars across years in university. My thesis was about development of artificial intelligence of a Japanese game called Shogi. I work at the private customs bureau as an IT also work at Teleperformance for Dell as an customer service for fixing electronic devices which means at the same time I was also doing IT works there. My best attributes are giving my all for the work I'm doing. I finish my job in time and do whatever it takes to make it right. I'm a very responsible person. I always tell to truth there is no need for lying. Time and money does not matter for me as long as I'm doing the job I like. I'm pretty good at English and I know a little of Japanese from the time I took class at university. As a programming and design languages I took Dr. Racket Python Java C C++ Assemble Mysql Django courses. Best Regards > |

Figure 3.3 shows the construction of matrices for the amount of information sharing that four OSN users perform on three profile items. First of all, GM is obtained from Table 3.4. Within the scope of this study, this matrix was created with the data of 5,389 real ProOSN users. Two different matrices are obtained using GM: PM (Privacy Policy Matrix) and GLM (Granularity Level Matrix) which are shown in Figure 3.3. From these matrices, PM is generated depending on whether or not these users share information. This matrix specifies the privacy policy of users based on their profile items. In addition, their granularity levels are assigned to GLM by making 1-dimensional clustering with dynamic programming (Wang & Song, 2011) for each attribute on GM. This matrix specifies 4 different levels of granularity on users' attributes: Zero Granularity (indicating no-share), Low Granularity, Medium Granularity, High Granularity levels.

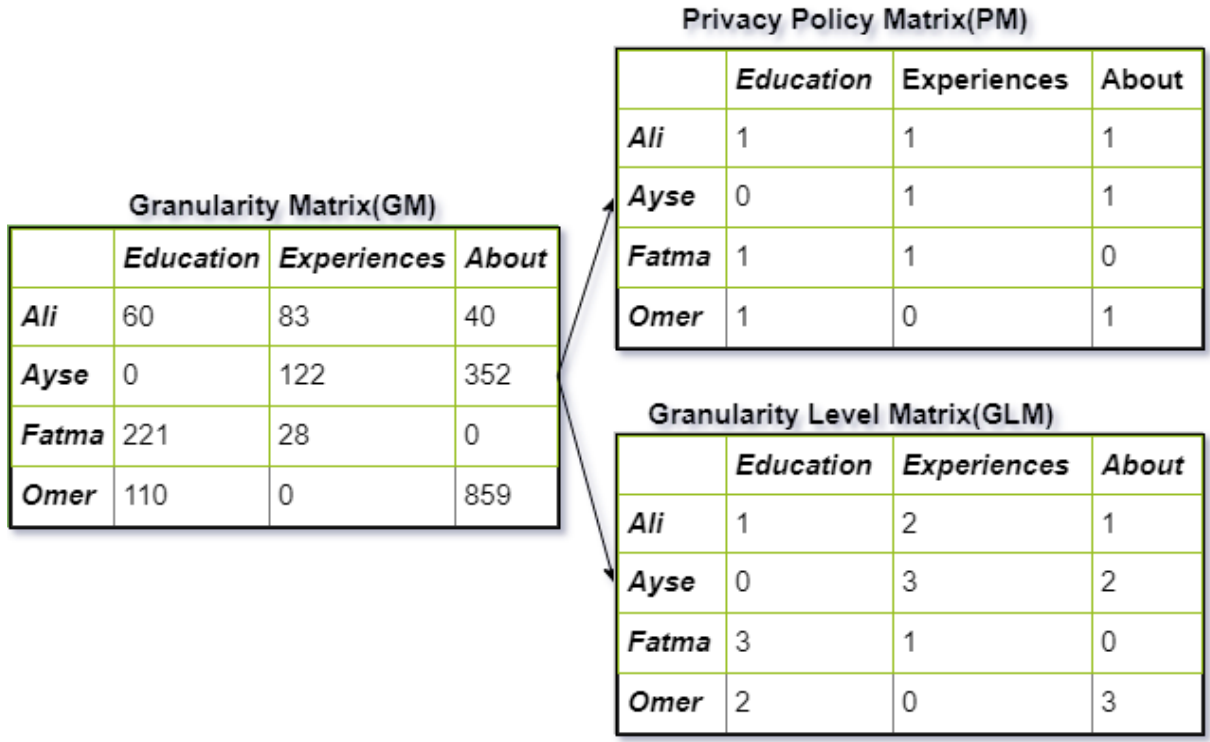

**Figure 3.3** Matrix Constructions

$$PSG^j = \sum_{i=1}^{n} \sum_{k=0}^{l} \beta_{ik} \; x \; G(i,j,k) \tag{3.12}$$

Equation 3.12 gives the general equation of the proposed scoring method. This scoring method is adopted from the polytomous problem settings of (Liu & Terzi, 2010). In general, users' privacy scores are the cumulative sum of the sensitivity and expected granularity. $\beta_{ik}$ denotes sensitivity value of profile item $i$ with respect to granularity level. $G(i,j,k)$ refers to the expected level of granularity for the level $k$ for item $i$ and user $j$. The calculation of the $G(i,j,k)$ value is shown in Equation 3.13. The $P_{ijk}$ value required for the calculation of $G(i,j,k)$ indicates the probability that user $j$ will share information on attribute $i$ at the $k$ level of granularity.

$$G(i,j,k) = P_{ijk} \times k \qquad (3.13)$$

As a granularity-based scoring model, we propose two different methods: PSGN, PSGI. These methods generally make the scoring using equations 3.12 and 3.13. However, the sensitivity and expected granularity level calculation approaches are different for these methods we proposed. While Section 3.3.1 provides information about PSGN, Section 3.3.2 explains PSGI model.

### 3.3.1 Naïve Based Granularity Model (PSGN)

This scoring method produces a probabilistic solution taking into account the users' sharing frequencies. We were inspired by the Naïve polytomous problem setting (Liu & Terzi, 2010). The assumptions adopted in this solution are also adapted for this method:

- Users and profile items are independent.
- The selection used by a small number of users is more sensitive.

$$\beta_{ik}^{Naive} = \frac{N - \sum_{j=1}^{N} I_{GLM_{i,j} \geq k}}{N} \qquad (3.14)$$

$$P_{ijk}^{Naive} = \frac{\sum_{j=1}^{N} I_{GLM_{i,j}=k}}{N} \times \frac{\sum_{i=1}^{n} I_{GLM_{i,j}=k}}{n} \qquad (3.15)$$

Based on all these assumptions, the equation of sensitivity value is given in Equation 3.14. In this equation, *N* gives the total number of users, while $GLM_{i,j}$ gives the level of granularity on *i*th attribute of the *j*th user in GLM. Function *I* gives the number of users using that granularity level according to the given granularity level. That is, the less preferred an item, the more sensitive it is accepted. Equation 3.15 gives the probabability of sharing an item by the user at the given granularity level. This equation is based on the assumption that users and attributes are independent. That is, the probability of a user *j* sharing the attribute *i* at the granularity level *k* is equal to the probability that user *i* is sharing at the *k* granularity level and that item *i*'s information sharing probability at granularity level *k* (See Equation 3.15).

### 3.3.2 IRT Based Granularity Model (PSGI)

The PSGI scoring method is derived from the GRM model (Samejima, 1997), an approach that measures the difficulty and ability of users of polytomous graded responses in Item Response Theory (Baker, 2004). In the GRM model, the most valuable answer is considered to be the most accurate. According to these graded responses, the difficulty of each response and the ability of the participants is defined. The focus of this study is as follows: The users' granularity levels are considered as polytomous forms of response, and the maximum value of 3 for each attribute is considered the most accurate answer. The ability values of the users produced by the GRM model and the level-based difficulty values for each attribute are input to the scoring function in Equation 3.12 to calculate the privacy scores of the users. In this calculation, unlike the PSGN model, the probability of sharing an item in Equation 3.16 according to the given granularity level is included. This probability calculation is used to compute the expected granularity level (See Equation 3.15).

$$P_{ijk}^{IRT} = \frac{1}{1 + e^{-\alpha_{ik}(\theta^j - \beta_{ik}^{IRT})}} \quad (3.16)$$

In addition, the R programming language with the LTM package was used for PSGI scoring model. The GRM method defined in the LTM package is given to GLM as a parameter, and this method calculates the sensitivity values, item discriminations, and user abilities of each granularity level of the items. Using these calculated values in Equation 3.16, the PSGI model calculates the probability that user *j* will share information for the *i* attribute at the *k* granularity level. Based on this probability, the privacy scores of the users are generated according to the general equation in Equation 3.12.

## 3.4 Evaluation Methods of Privacy Risk Scoring Models

Some statistical methods are commonly used to evaluate the scoring models. Two methods are commonly employed: $X^2$ Goodness of Fit Test, Correlation Analysis. We detailed these evaluation methods in this section.

### 3.4.1 Chi-Square Goodness of Fit Test

Chi-Squared Test determines how statistically significant the expected value and observed value (Wikimedia Foundation, Inc., 2019). This test was also used to compare privacy scoring models in terms of fitting to the data (Liu & Terzi, 2010).

First, users are clustered into *k* groups according to the estimated attitude values. Then, the significance of the difference between expected and observed values is tested with respect to degree of freedom. We took the degrees of freedom to be (*K* − *1*) for PSN and (*K* − *2*) for PSI where *K* is the number of groups. Equation 3.17 gives this Chi-Square Test formula (Liu & Terzi, 2010). Used terms and their descriptions are as follows:

- $F_g$ : Group g
- $f_g$: Number of users in $F_g$
- $p_{ig}^l$ : Observed probability of users in $F_g$ that share item *i*
- $p_{ig}$ : Expected probability of users in $F_g$ that share item *i*
- $q_{ig}^l$ : Not observed probability of users in $F_g$ that share item *i*
- $q_{ig}$ : Not expected probability of users in $F_g$ that share item *i*

$$X^2 = \sum_{g=1}^{K} \left( \frac{(f_g p_{ig}^l - f_g p_{ig})^2}{f_g p_{ig}} + \frac{(f_g q_{ig}^l - f_g q_{ig})^2}{f_g q_{ig}} \right) \quad (3.17)$$

**3.4.2 Correlation Analysis**

In statistics, statistical relationship between any two variables is called *correlation*. Correlation also gives the strength and direction of the relationship between two variables (Wikipedia Inc. , 2019). It is the ratio of the product of the covariance of two distributions to the product of standart deviations. Equation 3.19 shows this formula.

$$r_{X,Y} = \frac{\sum_{i=1}^{n}(X_i - \mu_X)(Y_i - \mu_Y)}{\sigma_X \sigma_Y} \tag{3.19}$$

# 4. RESULTS AND DISCUSSIONS

This chapter includes the results and evaluations of the experiments performed on the collected data. First, the statistical properties of the collected dataset are analyzed in Section 4.1. Then, the preferences made during data modelling of the collected data are explained with technical reasons in Section 4.2. Section 4.3 gives detailed experiments on existing and proposed privacy scoring models and finally statistical correlation between network centrality and policy based privacy scoring models is observed to understand privacy attitude of ProOSN users in Section 4.4.

## 4.1. Dataset Statistics

This section provides quantitative information about the structure and content of the collected data. In addition, these values are evaluated statistically.

Due to restrictions imposed by LinkedIn OSN Provider at the time of data collection, a seed user account and connections of the account, as well as connections of these connections can be accessed in user space. That is, access to the LinkedIn OSN is limited upto the second-degree connections of the seed account. From this seed account, 109 first degree and 26,208 second degree user IDs (in URL) were obtained. The number of users accessed depending on the distance (hop count) from the seed account is listed in Table 4.1. Due to LinkedIn OSN privacy policy, further degree connections are not available. This problem can be overcome by various methods such as initiating the crawler many through, seed with multiple different seed accounts. However, this solution could not be applied because there does not exist any secondary accounts we could utilize towards this end.

**Table 4.1** Number of Users Statistics with respect to hop count

| Hop Count | 0 | 1 | 2 |
|---|---|---|---|
| **Number of Users** | 1 | 109 | 26,208 |
| **Number of Profilling Users** | 1 | 109 | 5289 |
| **Number of Users Whose Connections are Profiled** | 1 | 44 | 0 |

It is clear that the structure of the graph model collected within the scope of this study is important for various social network analysis approaches. Figure 4.1 plots the structure of the collected social sub-network. Nodes on the network endpoints indicate users whose connections cannot be visited. In addition, users at the center of the network represent central users who have more connections with other users within the network.

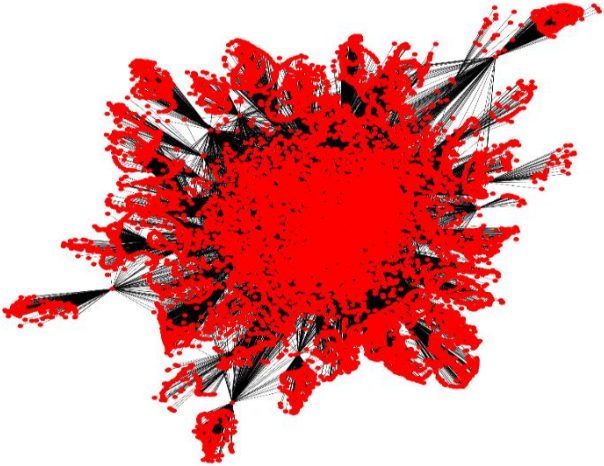

**Figure 4.1** Graph Structure of the Collected Dataset

**Table 4.2** Social Network Metrics of the Collected Dataset

| **Dataset Graph Properties** | **Metric Values** |
|:---:|:---:|
| Number of Nodes | 5,389 |
| Number of Edges | 40,009 |
| Average Clustering Coefficient | 0.0722 |
| Network Diameter | 4 |
| Average Path Length | 2.34 |

Social network metrics of the collected graph dataset are shown in Table 4.2. The number of nodes refers to the number of unique users. The number of edges gives the total number of connections between these users. LinkedIn connections are undirected. That is, if a user *A* is connected to user *B*, it means that user *B* is also connected to user *A*. The grouping coefficient gives the ratio of how well the connections of a user's connections are completed. The average of all users gives the average grouping coefficient. The average coefficient of data collected in this study is very low: 0.0722. The reason for this can be interpreted as poor connections between users' connections. Also, the diameter of the network is 4. So, the distance between the two furthest users on the network is 4. In addition, the average path length was observed to be 2.34. These graph properties indicate that our OSN data is sparse, and social connectivity between users is weak. Consequently, central users of the LinkedIn OSN are vital in reaching a large number of users in the data collection phase.

**4.2 Modelling Preferences**

In this section, the rationale behind our data models is presented. We also discuss the efficiency issues related to OSN data analytics for the different database models that were considered.

The collected data was modelled with two types of database: Relational Database (RDB) and Graph Database (GDB) model. For offline data analysis studies, the data were stored with these two common database models. Figure 4.2 shows RDB model of the collected dataset and Figure 4.3 gives GDB model of the collected data. The LinkedIn Graph Database Model is presented in Figure 4.3. The figure is taken from Neo4j database after entering the records of several accounts and appears much simpler that it actually is because the node and edge properties are excluded.

In terms of social dimension, it isn't efficient to analyze OSN data -except for a limited range of queries- even though it is quite easy to store it in Relational Database systems. Below are a few examples of queries that cannot be written on RDB or cannot be answered efficiently. Our first example is the following query: "*list all accounts* Y *that are friends of any degree with account* X." This query fetches all friends $Y$ of $X$ and requires joining the friendship relationship relation infinitely many times in relational model under RDB – an obviously inefficient strategy. Another example is to determine whether the collected OSN could alternatively be generated from another seed account *Y*. This analysis is carried out quite easily and naturally with the following path query in the Cypher language of Neo4J: "*MATCH (*X*: User {name: 'Ayse Fatma'}) MATCH (*Y*: User) MERGE (*X*)-[Friend*]->(*Y*) RETURN* Y". This path query takes about 10 minutes to execute in Neo4j. However, the corresponding task requires many steps to be taken over a relational database. Friends of account *X* can be found using friendship table in RDB systems. If this table is merged with itself, *X*'s friends can be located. However, it is not possible to write a query based on a random length path in RDB system languages such as SQL. There is only one solution to develop and run a DFS (Depth First Search) algorithm starting from an account *X* using a high tier structural query language such as PL/SQL. We implemented this complex PL/SQL query in our RDB model and this simple experiment took about 26 hours for a single node *X* to return a result.

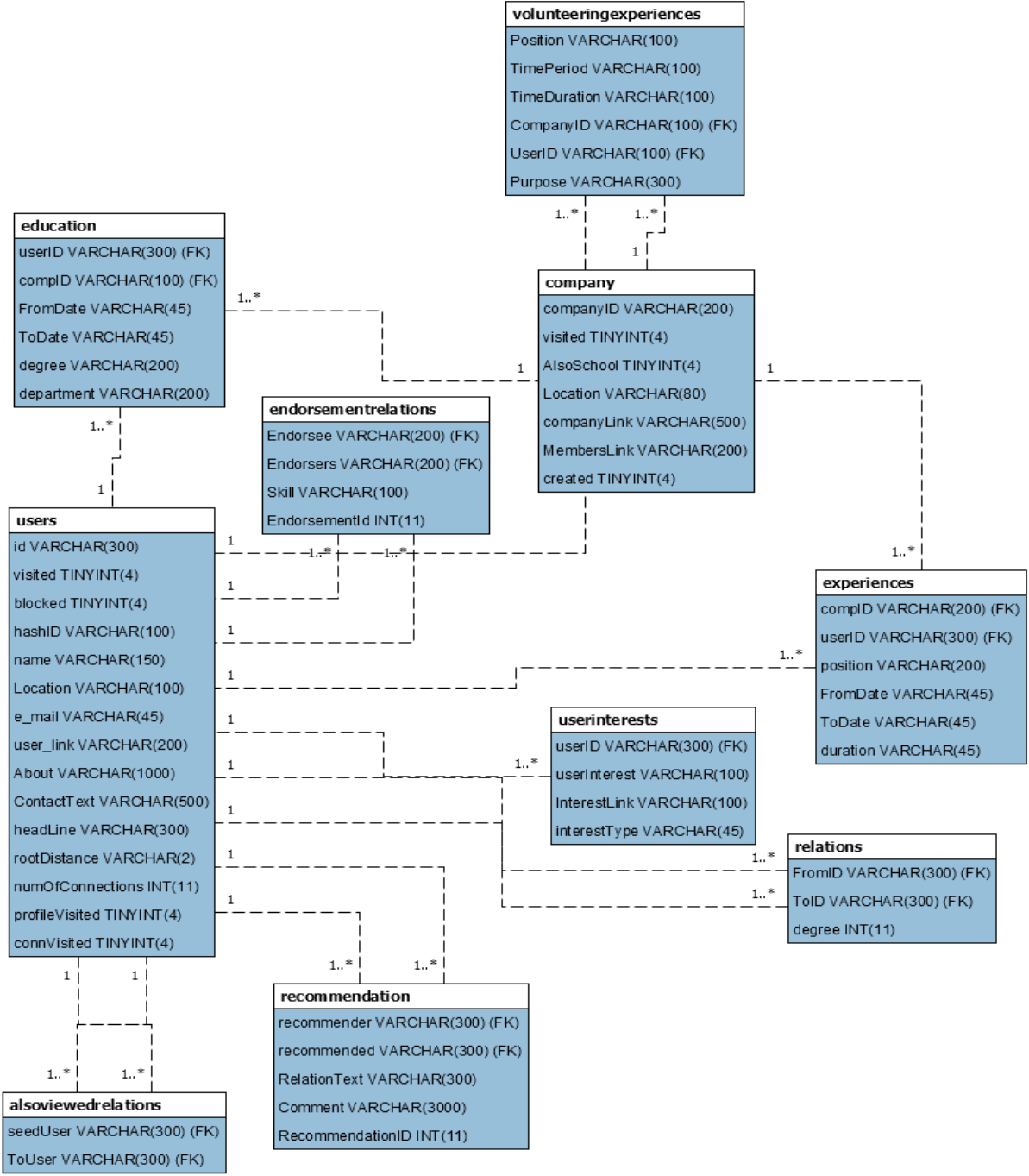

**Figure 4.2** Relational Database Modelling of the Collected Dataset (From MYSQL)

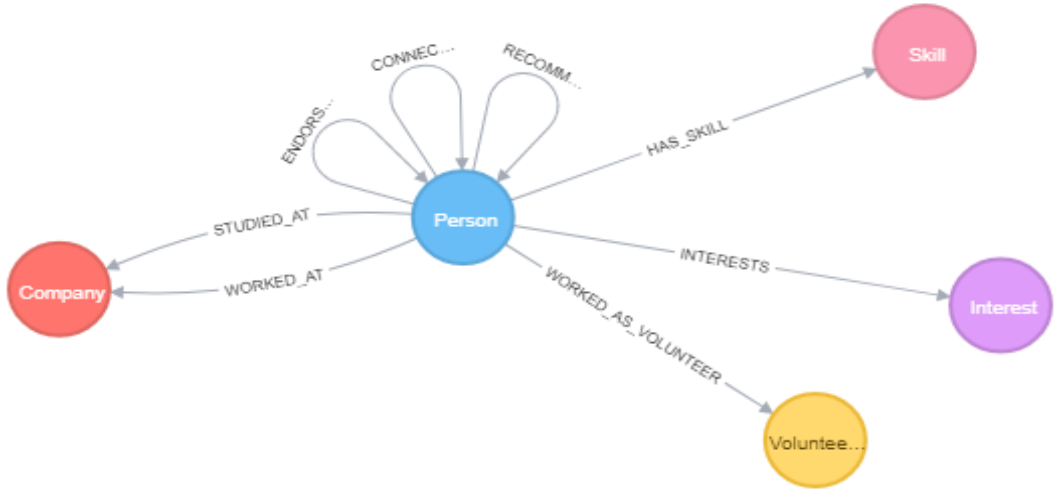

**Figure 4.3** Graph Database Modelling of the Collected Dataset (From Neo4j)

It is necessary to merge all relationship tables (friendship, kinship, liking, etc.) on RDB systems. In addition to being difficult to write this type of queries correctly in SQL, it will not be possible to get the answer with RDB Management Systems in terms of query cost. On the other hand, this type of queries is written quite simple and can be answered very quickly by the index-free adjacency principle (Robinson, Ian, Webber, & Eifrem, 2015). According to this principle, for each edge *R* defined from node *X* to node *Y*, an *R* pointer is defined from node *X* to *Y*. Thus, even if "*X-R->Y* "relationship is not indexed, *X* to *Y* query can be easily answered. Neo4j's Cypher Query is as follows: "*MATCH (X: User {name: 'Ayse Fatma}) MATCH (Y:User) MERGE (X)-[R]->(Y) RETURN X, R, Y*" .

### 4.2. Evaluations of the Existing and Proposed Privacy Scoring Models

In this section, the existing and our proposed privacy scoring approaches are implemented on the collected dataset and the results are analyzed comparatively.

We first compared two privacy policy-based scoring models (PSN, PSI) in terms of the sensitivity values of the profile items listed in Table 3.2. The results are depicted in Figure 4.4. Notice that the item sensitivity values for $\beta_i^{Naive}$ of Equation 3.2 and $\beta_i^{IRT}$ of Equation 3.5 are quite distant. Recall that the sensitivity of an item determines the extent to which the item increases the privacy risk of the user sharing it publicly. For example, intuitively, the phone number is expected to be more sensitive than education. Both models respect this anticipation. However, notice also that in the Naive model, the two items are only marginally different in item sensitivity. The IRT model's performance in sensitivity measurement is much better because this model captures each user's sharing attitudes into account.

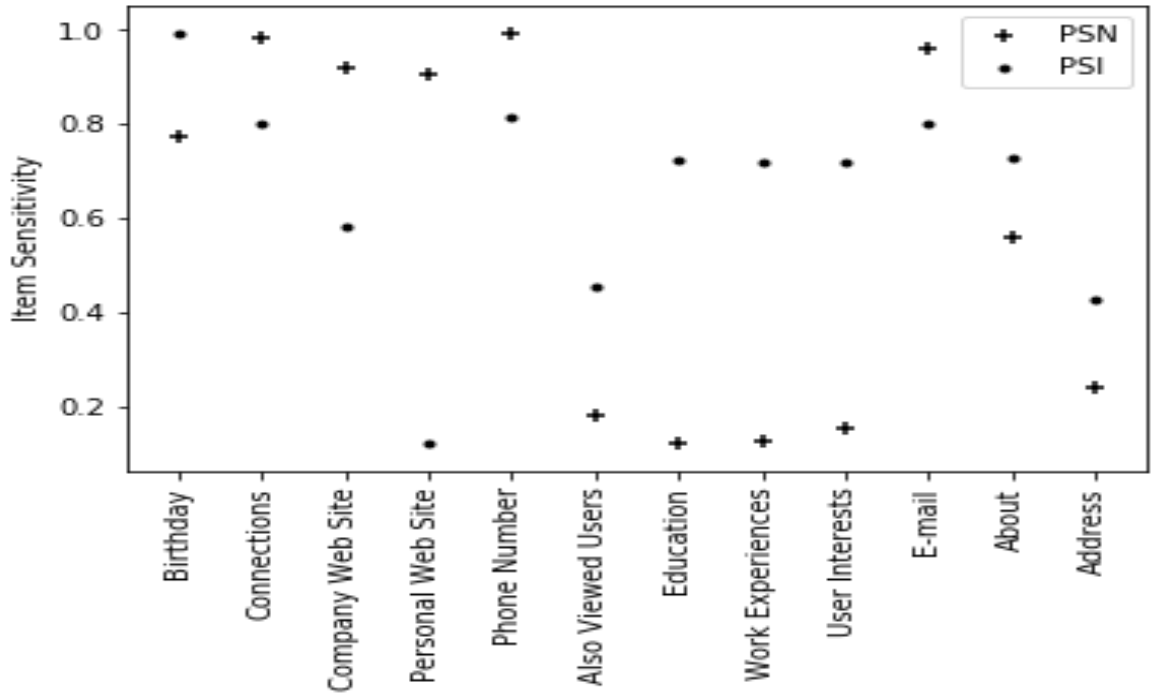

**Figure 4.4** Item Sensitivity Per Item

Table 4.3 involves goodness of fit tests for privacy policy-based privacy scoring models which are PSN, PSI. We relied on chi-square hypothesis testing to determine which of the models (e.g., PSN and PSI) better fits the underlying response or policy matrix. We took the degrees of freedom to be $(K-1)$ for PSN and $(K-2)$ for PSI where $K$ is the number of groups. Under conditional independence assumption over items, we repeated the goodness of fit tests for each of the 12 items listed in Table 3.2. Table 4.3 summarizes the results. Regardless of the number of groups (please refer to (Liu & Terzi, 2010) for details of grouping and group invariance properties of IRT), more PSI hypothesis were accepted than PSN hypothesis. This indicates that PSI fits the matrix better than PSN.

**Table 4.3** Goodness of Fit Test Results for Policy Based Privacy Scoring Models

| Number of groups(K) | PSN | PSI |
|---|---|---|
| 3 | 5 | **7** |
| 4 | 5 | **7** |
| 6 | 5 | **7** |
| 8 | 5 | **7** |
| 10 | 2 | **10** |
| 12 | 3 | **9** |
| 14 | 3 | **9** |

Next experiments focus on our proposed granularity-based scoring models (PSGN, PSGI). Various analysis was performed to compare these models.

Figure 4.5 shows the distribution of user's profile information granularity. It is observed that there is significant variance between the granularity values and the grouping approach with respect to granularity values may be suitable. In Figure 4.5, mean and standart deviation values of the granularity distributions for each profile item are shown. Our approach is based on the amount of information that users share. In order to analyze this basis experimentally, statistical values of the amount of information were observed for each attribute. Figure 4.5 shows the average amount of information and standard deviation values for each attribute. It is observed that the standard deviations of some profile items, such as phone numbers and birthdays, are very low or zero. The reason for this is related to the nature of the item. For example, the amount of information entered for the birthday attribute is generally fixed length. However, it is seen that standard deviation values are high in the items such as education, work experience and address. The scoring approach proposed in this study suggests assigning different scores to users who share the same item in different details. For example, it is unrealistic that a user sharing 15 bytes of information for training information and a user sharing 300 bytes of information have the same privacy score. This experiment also states that proposed method is necessary to capture privacy risk of OSN users because of almost profile items have high standart deviations.

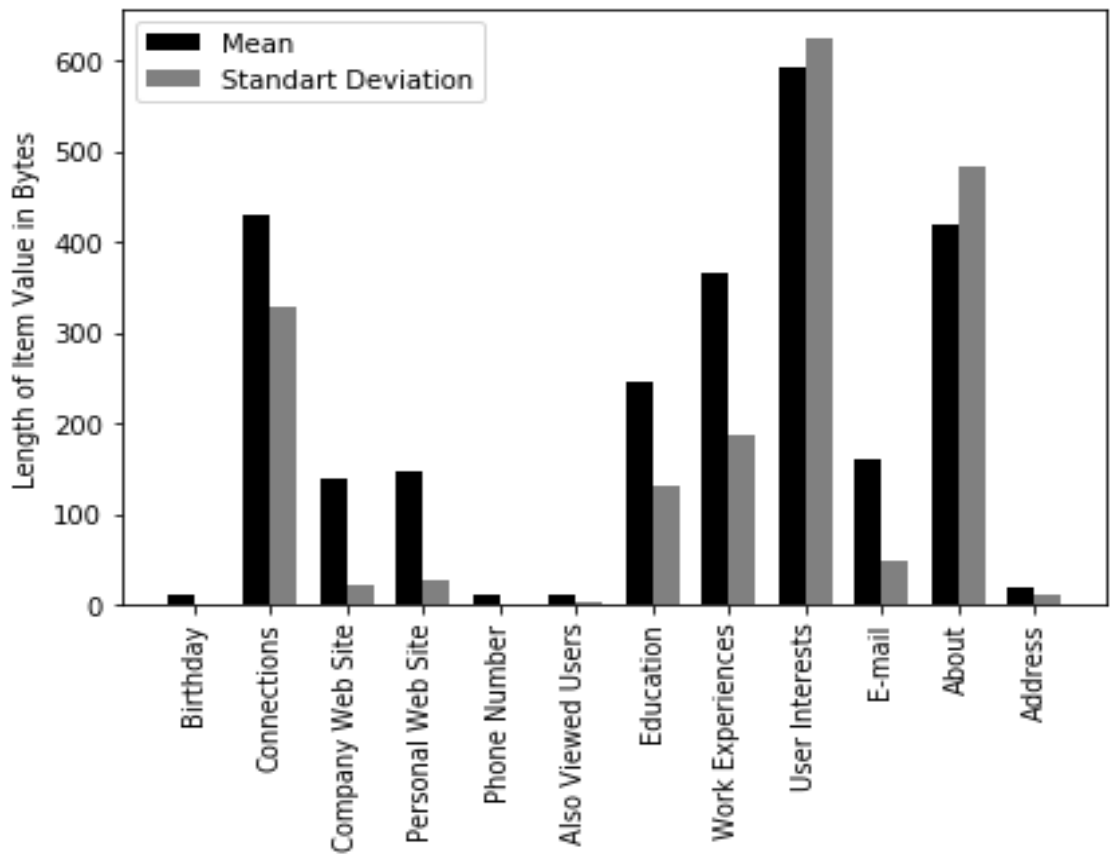

**Figure 4.5** Statistical Properties of Item Granularities

**Table 4.4** Goodness of Fit Test Results for Granularity Based Scoring Models

| Number of Groups | PSGN | | | PSGI | | |
|:---:|:---:|:---:|:---:|:---:|:---:|:---:|
| K | R1 | R2 | R3 | R1 | R2 | R3 |
| 3 | 9 | 6 | 7 | 3 | 6 | 5 |
| 4 | 5 | 6 | 6 | 7 | 6 | 6 |
| 6 | 3 | 4 | 2 | 9 | 8 | 9 |
| 8 | 3 | 2 | 2 | 9 | 9 | 9 |
| 10 | 2 | 1 | 2 | 10 | 11 | 9 |
| 12 | 1 | 1 | 2 | 11 | 11 | 9 |
| 14 | 1 | 1 | 2 | 11 | 11 | 9 |

Table 4.4 shows which model best fitted to data for the proposed granularity-based models (PSGI, PSGN). According to test results in Table 4.4, we clearly say that PSGI fits GLM better than PSGN. In PSGI; by grouping number increases, it is observed that performed better fitting on the granularity matrix. So PSGI model can be selected among granularity-based privacy scoring models.

In Figure 4.6, sensitivity values produced by granularity based PSGN and PSGI models are compared. As we know, these models produce different values for each granularity level $k$. For this experiment, the values produced for all granularity levels are compared for each profile item. Both models look almost similar in relative sensitivity magnitudes at high level granularity. In general, both models give approximately the same sensitivity value for high level granularity ($k=3$). But for other levels, items with quite differences are observed such as education, experiences, user interests at low level ($k=1$). We think this is due to difference in scoring approaches of the models. We know that PSI model takes into account the ability of users while PSN model considers the frequency of sharing only.

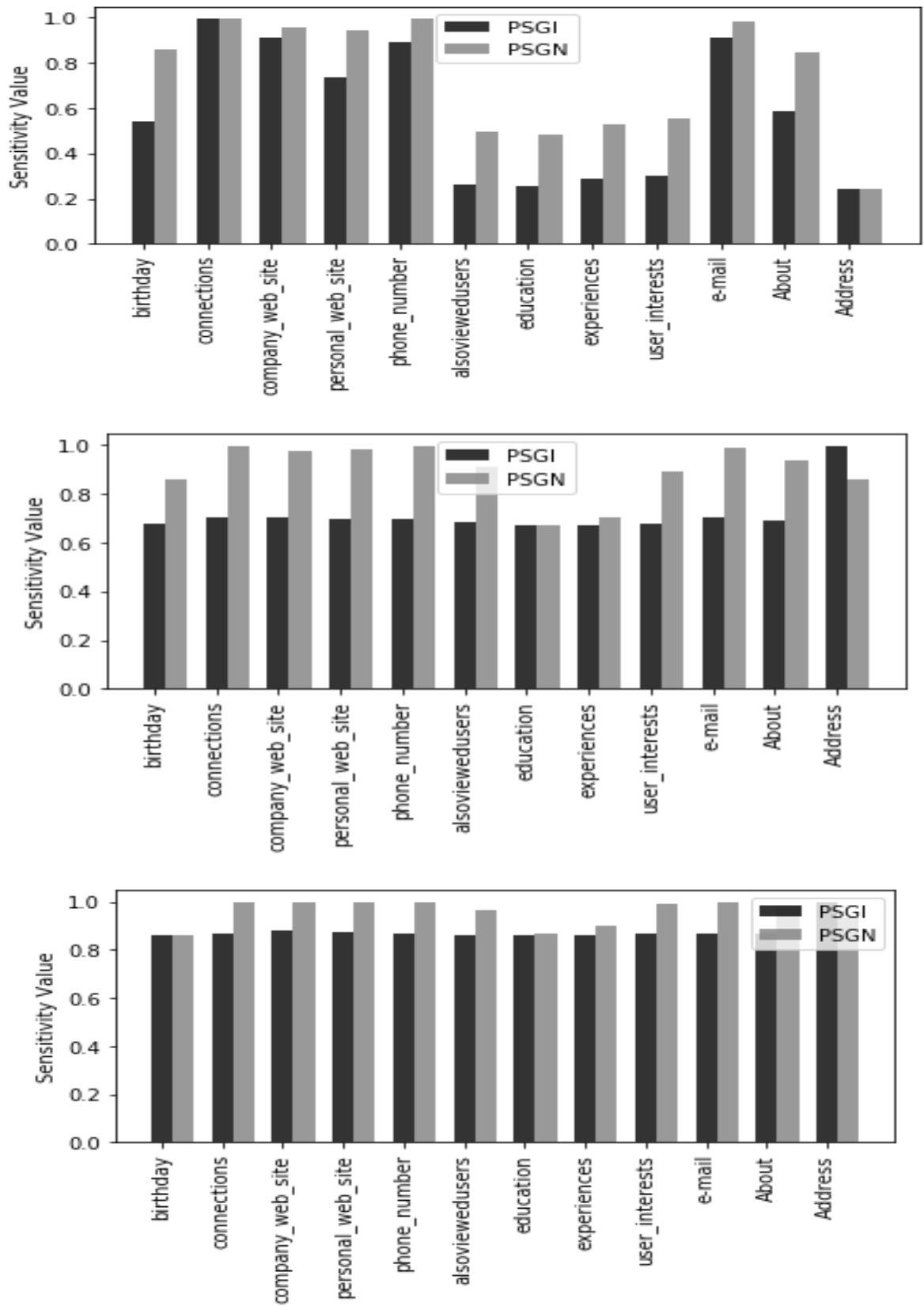

**Figure 4.6** Sensitivity Comparison by Granularity Models: (Top) k = 1, (middle) k = 2, (Bottom) k = 3

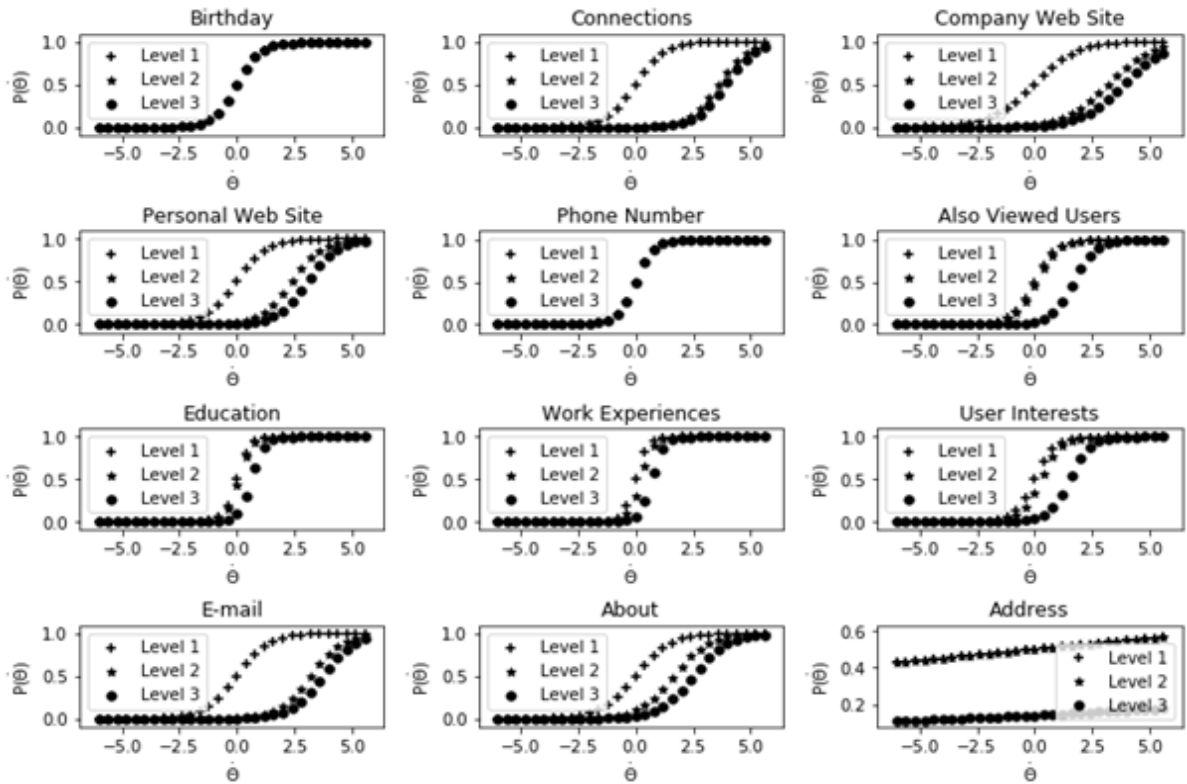

**Figure 4.7** Item Characteristic Curves with respect to Granularity Level

Experiment shown in Figure 4.7 gives Item Characteristic Curves (Baker, 2004) which is probability value of correctly response for each granularity level with respect to user sharing attitudes which denotes $\theta$. As the level of granularity increases, it is observed that the difficulty values on all profile items are increase. This observation shows that the model is consistent and provides following our intuitive corollary: more information is more difficult to share. In remainder of this experiment, the items are evaluated in terms of difficulty based on user abilities ($\theta$ in Figure 4.7). We observed some findings about ProOSN profile items as follows:

- *Address* item appears to be less sensitive at low (Level 1) and medium (Level 2) granularity. However, it is very sensitive for high granularity (Level 3). Because, while theta value increases, the probability value increases slightly.
- Since the amount of information sharing for *phone number* and *birthday* items is equal lengths. So, these items are modelled as dichotomous or binary. That is either shared or not shared. Therefore, ICCs of these items have a single curve. It is observed that they have similar difficulty values according to the curves. But, the steeper of the curve of phone number item indicate that this attribute is discriminable. This means that the probability value will increase faster depending on the abilities of users ($\theta$ value in Figure 4.7).

- *Education* and *Work Experiences* items seem to have low sensitivity or difficulty values. The fact that the curves are close to each other shows that there is no obvious difference in sensitivity between granularity levels. However, sensitivity difference among granularity levels has been observed for users with obvious differences between granularity level curves such as *Connections, E-mail*. It has observed that having more connections and longer e-mails are more sensitive choices than other options or granularity levels.

**Table 4.5** Statistically Correlation Between Privacy Scoring Models

|  |  | *Policy Based* | | *Network Based* | | *Our Granularity Based* | |
|---|---|---|---|---|---|---|---|
|  |  | *PSN* | *PSI* | *PSC* | *PSNA* | *PSGN* | *PSGI* |
| *Policy Based* | *PSN* | 1 | 0.31 | 0.28 | 0.30 | 0.34 | 0.26 |
|  | *PSI* | 0.31 | 1 | 0.34 | 0.37 | **0.20** | **0.22** |
| *Network Based* | *PSC* | 0.28 | 0.34 | 1 | 0.81 | 0.067 | 0.045 |
|  | *PSNA* | 0.30 | 0.37 | 0.81 | 1 | 0.003 | -0.007 |
| *Our Granularity Based* | *PSGN* | 0.34 | **0.20** | 0.067 | 0.003 | 1 | 0.90 |
|  | *PSGI* | 0.26 | **0.22** | 0.045 | -0.007 | 0.90 | 1 |

Last experiment investigates statistical correlation between existing and proposed privacy scoring models. The correlation values of six scoring methods of 3 different scoring models are listed in Table 4.6. Although these values do not necessarily give causality, they can provide important clues about statistical relationship between these scoring methods and models. As a result of our observation, the following findings are obtained:

- There is a positive but not very strong relationship between Policy-based scoring models (PSN and PSI): 0.31. The reason why this positive relationship is not very strong is related to scoring approaches. While the PSN method calculates users' policy frequencies independently for each attribute, the PSI method takes into account the policies that users have over other attributes by proposing a more comprehensive approach. Already in Table 4.3, the test of goodness of fit clearly shows that the PSI method fits the data better.

- A Strongly positive correlation was observed between Network Based Privacy Scoring Models (PSC, PSNA). As we know, PSC refers to users' network position without privacy policy, while PSNA references both privacy policy and network position of the users. It is clearly that PSNA overestimates the network positions.

- A strongly positive correlation was observed between the granularity-based models (PSGI and PSGN):0.90. This means that both models give statistically similar scores. But PSGI model fits the data better according to Table 4.4. Thus, the PSGI model is accepted as best granularity-based model. We observed that IRT based models (PSI, PSGI) seem to fit data better. These observations prove that IRT based models have more group invariance property (Baker, 2004) (Liu & Terzi, 2010).

- Our best proposed model –PSGI- doesn't take into account topological properties of users such as network centrality. Therefore, it would be more appropriate to evaluate with policy-based models (PSN, PSI). For example, because the address item is shared by most users, the sensitivity for the Policy-based models is considered low, while the Granularity-based models determine the sensitivity value based on how much granularity level the user shares the address item. If this user shares the address item at a high granularity level, a higher sensitivity value is assigned. Moreover, we observed a suitable correlation between policy and granularity-based models that was too low to be considered as a duplicate and too high to be independent (*0.22*). We think that our best granularity-based scoring model (PSGI) contributes a new comprehensive dimension or factor to used scoring models.

- Some interesting results were also obtained according to this experiment. While a positive relationship was observed between Network-based models and Policy-based models, it was observed that there was a statistically low relationship with Granularity-based models. This result shows us that central users tend to share sensitive items but share these sensitive attributes in less detail or vice versa.

To summarize, privacy attitudes of the users appear to depend on their granularity level on profile items in the OSN. In addition, these experiments show that probabilistic scoring models (PSI, PSGI) derived from the Item Response Thory (IRT) fit real world data better than naïve approaches (PSN, PSGN).

### 4.3. Investigating Correlation Between Topological Properties and Privacy Scores

This section investigates the relationships between privacy policy-based privacy scores which are PSN and PSI models and network position of ProOSN users.

First; the effect of the damping factor value which is hyperparameter for Page Rank Centrality (PRC) is observed. Figure 4.7 shows the correlation between PSN and PSI models against the Page Rank algorithm in increasing damping factor. We observe that regardless of the damping factor, both Naive and IRT based privacy scores are positively correlated with Page Rank scores. This indicate that damping factor can be ignored that is default damping factor can use. Default damping factor is 0.85. In this study, this value is used as damping factor.

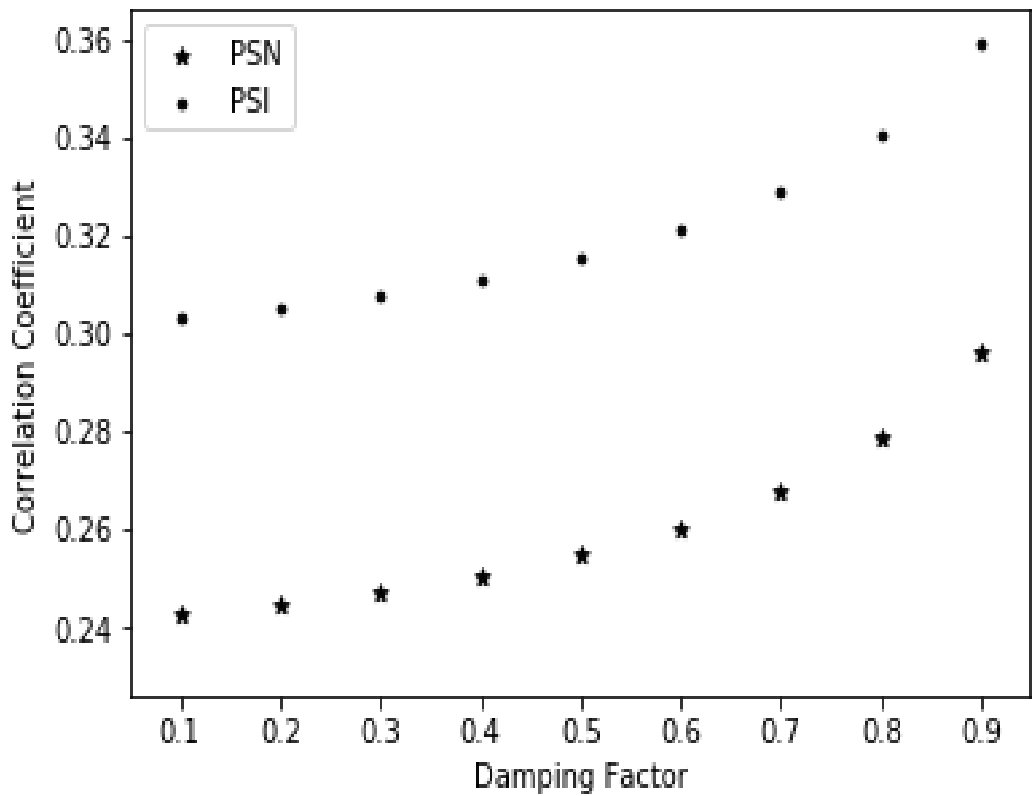

**Figure 4.8** Page Rank Damping Factor Effect

The next experiment analyzes the relationship between network centrality value and policy-based privacy scores. In fact, this experiment aims to answer the following research question:

What is the position of users on the network who share more sensitive items of users, i.e high risk of privacy? In order to answer this research question, the correlation values between privacy policy based intrinsic privacy scores generated by users' profile items sharing policies and network centralities are shown in Table 4.7. Four commonly used network centrality methods are implemented as follows: Page Rank Centrality (Brin & Page, 1998), Eigenvector Centrality (Ruhnau, 2000), Closeness Centrality (Brandes, 2005, February) and Betweenness Centrality (Freeman, 1977). These four centrality methods use different approaches. This is why the correlation values in the table are different. The lowest correlation value is observed in the BC method. This means that less statistical relationships are observed with Privacy policy-based privacy scores of the users that are central for information flow across the network.

**Table 4.6** Correlation between Policy Based Privacy Scores and Centralities

|  |  | *Policy Based Scoring Models* | |
|---|---|---|---|
|  | ***Correlation*** | *PSN* | *PSI* |
| *Centrality Methods* | *PRC* | 0.2865 | 0.3487 |
|  | *EVC* | 0.3066 | 0.2932 |
|  | *CC* | 0.3695 | 0.2965 |
|  | *BC* | 0.1871 | 0.2451 |

Policy-based scoring models calculate users' sensitive information based on their approach to sharing information. When the correlation values expressed in Table 4.6 were examined, a positive relationship was observed for all centrality scores. According to studies in the literature (Pensa & Blasi, 2016) (Liu & Terzi, 2010), this relationship is relatively high. This is thought to be due to the fact that central users share more sensitive items. While the studies in the literature were conducted over PerOSN users and their social network, this study was conducted on ProOSN users and their social network. We think that different results are related to the sharing attitude of these OSN users. In other words, OSN users may have different sharing attitude in different OSN platforms such as Facebook as PerOSN, LinkedIn as ProOSN.

As a result of this observations, privacy policy attitudes of ProOSN users appear to depend on their centrality in the network. It is possible that influential or central users are under higher risk because they are sharing their sensitive items aggressively. Unfortunately, based solely on correlation, drawing conclusions of causality between the centrality and the policy-based privacy score of a user would not be correct.

# 5. CONCLUSIONS

Existing work on privacy scoring over OSNs worked with survey data that has been criticized for survey responses being emotional or biased. In this work, we collected real world OSN data from a ProOSN- LinkedIn OSN. Our empirical results over the profiles of 5,389 distinct users reveal that, (i) privacy score of a real-world OSN user can be modelled accurately using IRT based models such as PSI and PSGI, (ii) this score is correlated with the network position of the users, (iii) unlike the claims of existing work, the correlation is relatively positive: This means, more central users tend to have higher privacy scores while less central users tend to have lower privacy scores and vice versa, (iv) granularity is a comprehensive factor for privacy scoring problem and we empirically show through extensive experiments that shared data granularity does contribute to privacy scoring significantly.

# 6. RECOMMENDATIONS

In future work, we plan to more closely investigate whether there is any causality between these two dimensions as well as incorporating additional factors into the scoring mechanism such as granularity of the shares and the sensitivity of the shared entry data over the entire OSN. On the other hand, existing evaluation of privacy scoring models based on statistical analysis such as goodness of fit test. In fact, users' how vulnerable to privacy attacks in real life is scored. We plan to evaluate users' privacy scores by creating real privacy attack scenarios in the future.

Wani, M., Agarwal, N., Jabin, S., & Hussai, S. (2018). Design and Implementation of iMacros-based Data Crawler for Behavioral Analysis of Facebook Users. *arXiv*.

We Are Social Ltd. (2018, May). *We Are Social - Digital Report 2018*. Retrieved from https://digitalreport.wearesocial.com/: https://digitalreport.wearesocial.com/

Wikimedia Foundation, Inc. (2018, May). *Relational database*. Retrieved from Wikipedia Web Site: https://en.wikipedia.org/wiki/Relational_database

Wikimedia Foundation, Inc. (2019, November). *ACID:wikipedia*. Retrieved from wikipedia web site: https://en.wikipedia.org/wiki/ACID

Wikimedia Foundation, Inc. (2019). *Chi-squared test - Wikipedia*. Retrieved from Wikimedia Web Site: https://en.wikipedia.org/wiki/Chi-squared_test

Wikipedia Inc. . (2019). *Correlation and dependence - Wikipedia*. Retrieved from Wikipedia: https://en.wikipedia.org/wiki/Correlation_and_dependence

Yang, Y. L. (2012, February). Stalking online: on user privacy in social networks. *In Proceedings of the second ACM conference on Data and Application Security and Privacy* (pp. pp. 37-48). ACM.